# Is SARS-CoV-2 a new Frankenstein monster virus?

**Mohammed A. El-Magd**

Department of Anatomy, Faculty of Veterinary Medicine, Kafrelsheikh University, Egypt.

\***Corresponding author:** Mohammed Abu El-Magd,

E-mail address: mohamed.abouelmagd@vet.kfs.edu.eg,

ORCID: https://orcid.org/0000-0002-3314-9202

**Abstract**

Severe acute respiratory syndrome coronavirus 2 (SARS-CoV-2), a novel virus of beta coronavirus genus originated in Wuhan, China in December 2019, resulted in the pandemic spread of Corona Virus Disease 2019 (COVID-19) worldwide. This genus also contains SARS-CoV (originated in China 2002-2003) and MERS-CoV (found in Saudi Arabia 2012). The nucleotide sequences of SARS-CoV-2 is closer to SARS-CoV (with approximately 80% identity) than to MERS-CoV. Despite these similarities, SARS-CoV-2 has two main features over the other coronaviruses. The first is the high contagious rate and the second is the immune response evasion. The higher transmission ability makes this virus quickly spread worldwide with a high mortality rate and more economic losses. This review will provide an overview of the current knowledge on the role of some viral and host cell factors in higher transmission and contagious spread of SARS-CoV-2.

**Keywords**: Furin site; Receptor binding motif; CD147 receptor; Pericyte; Extracellular vesicle.

**Introduction**

Although no current accurate available data explain SARS-CoV-2 higher transmission and contagious spread, herein I speculated about 5 possible factors: 1) furin cleavage site in the spike protein (SP), 2) mutated receptor binding motif (RBM) of SP, 3) new receptor (CD147), 4) new target (pericytes), and 5) new entry route (extracellular vesicles, EV). The first 2 are virus-related factors and the remaining are host-related factors. Away from the conspiracy theory, I consider these 5 factors as pieces that were assembled by an unknown hand to create SARS-CoV-2 similar to the creation of Frankenstein monster (Fig.1).

**Furin site piece**

Before binding with angiotensin-converting enzyme 2 (ACE2) receptor, SP of SARS-CoV-2 should be cleaved by protease enzymes into the S1 domain (responsible for ACE2 binding) and the S2 domain (responsible for virus-cell membrane fusion) (Hoffmann et al., 2020). Three host protease enzymes are involved in S priming, including furin, trypsin, and transmembrane protease serine 2 (TMPRSS2). Unlike SARS-CoV SP, SARS-CoV-2 SP can be cleaved by all these enzymes (Coutard et al., 2020; Hoffmann et al., 2020; Li et al., 2020; Yuanchen et al., 2020). Another protease, cathepsin B/L (CTSB/L) also participates in SARS-CoV-2 entry (via endosomal pathway) especially in the upper respiratory tract (Sungnak et al., 2020). Interestingly, furin-mediated cleavage could provide a 100-fold higher efficient infection than other protease enzymes do (Matsuyama et al., 2005) and so this can explain the higher affinity for SARS-CoV-2 to enter cells relative to SARS-CoV (Li et al., 2020; Yuanchen et al., 2020). Notably, S protein of the infectious bronchitis virus (IBVs) Beaudette strain also contains furin cleavage site, which correlates with higher infection ability of IBV-Beaudette as compared to other IBV strains (Yamada and Liu, 2009). Additionally, the profuse expression of ACE2, furin, and TMPRSS2 in the lining epithelia of the upper respiratory tract enables SARS-CoV-2 replication in the upper respiratory tract before shedding to the lung and suggests its enhanced transmission (Yuanchen et al., 2020). The wide distribution of furin in most tissues of the body enhances the SARS-CoV-2 replication in non-respiratory organs such as intestine, kidney, and heart which could also explain the multiple symptoms of COVID-19 and multi-organs dysfunction in late stages (Hua et al., 2020). The perfect insertion of the furin cleavage site (9 inserted nucleotides coding 3 amino acids) in receptor binding domain of SP gene (at S1/S2 interface) makes scientists think in laboratory synthesis theory of SARS-CoV-2 (Coutard et al., 2020) but to date no scientific proof for this theory.

**RBM piece**

Some non-synonymous single nucleotide polymorphisms (SNPs) changed some amino acids and led to conformational changes in RBD of SP in a way that makes its binding with ACE2 very strong (Hua et al., 2020; Shang et al., 2020). Studying the crystal structure of RBD of the two SARS viruses using cryo-electron microscopy revealed a very stronger binding between SARS-CoV-2 and ACE2 receptors than SARS-CoV does (Wrapp et al., 2020). This binding also showed a clear conformational change in the main loop of the complex binding interface. Some researchers accept this theory that argues for the natural origin of the virus due to selective/adaptive mutations from a wild host (bat) to humans. However, this stronger receptor binding is not enough to fully clarify the more contagious rate.

**CD147 piece**

CD147 (also known as Basigin) is a transmembrane glycoprotein that acts as an extracellular matrix metalloproteinase (MMP) inducer. CD147 can act as a receptor, alternative to ACE2, for SP of SARS-CoV (Chen et al., 2005) and SARS-CoV-2 (Wang et al., 2020b) to enter the host cells. The anti-CD147 antibody Meplazumab has the potential to block the binding of SP and CD147 and arrest cell entry, which offers a potential opportunity for developing effective treatment or vaccine targeting CD147 receptor (Wang et al., 2020b). This monoclonal antibody is now using under a clinical trial in China (ClinicalTrials.gov Identifier: NCT04275245). Interestingly, CD147 was also abundantly expressed in RBCs and act as a receptor for Plasmodium falciparum that causes Malaria in humans (Crosnier et al., 2011). However, it is unknown whether SARS-CoV-2 can also bind to this receptor on RBCs. To date, no proof for infection of RBCs by SARS-CoV-2 or even other related coronaviruses. This important topic could be addressed in the next investigations. Only, one *in silico* study found that three non-structural proteins (orf1ab, ORF10, and ORF3a) can compete with iron to bind with porphyrin (Wenzhong and Hualan, 2020). However, this binding was not confirmed yet by in vivo or in vitro studies.

Diabetic patients have a higher expression of CD147 and its downstream target MMP1 and MMP9 (Ulrich and Pillat, 2020). This may explain why diabetic patients constitute a high-risk group for COVID-10 morbidity, but further studies are required to validate this possibility. The potential relive effect of azithromycin against COVID-19 could be attributed to the blocking of CD147 receptor (Ulrich and Pillat, 2020), however, this could first be experimentally confirmed. CD147 is extensively expressed in brain pericytes and its ligand cyclophilin A (CypA) activates this expression (Pan et al., 2020). On the other hand, cyclosporine A, as a CypA inhibitor, could inhibit the binding of CypA with CD147 in pericytes (Pan et al., 2020). Intriguingly, the potent immunosuppressive cyclosporine A drug can be used to reduce cytokine storm induced by SARS-CoV-2 and probably to inhibit CD147 expression, but its use is restricted due to its adverse effects (Cure et al., 2020)

**Pericyte piece**

Before corpus dissection by Italian scientists, all our attention was paid toward the severe acute lung injury associated with COVID-19. However, after the post-mortem examination (Magro et al., 2020), most physicians and scientists focused on cardiovascular morbidities and started to include antithrombotic drugs in the treatment protocol. Researchers looked for cellular components invaded by SARS-CoV-2 through determination which vascular cell expresses ACE2. There is a big debate regarding the expression of ACE2 by endothelial cells. Some studies confirmed this expression (Muus et al., 2020) and others excluded this expression and suggested the presence of contamination by pericytes (He et al., 2020). Unlike endothelial cells, pericytes (perivascular cells wrapped the endothelial cells and enclosed by the same basement membrane) expressed the ACE2 receptor. In the COVID-19-pericyte hypothesis postulated by He et al. (2020), targeting pericytes, instead of endothelial cells, by SARS-CoV-2 is the main cause for multiple thromboses which subsequently lead to death in high-risk people including diabetic, hypertensive, and obese. According to this theory, this high-risk group has a defective endothelial barrier (damaged junction complex between endothelial cells) which permits SARS-CoV-2 to reach the peripherally located pericytes and binds to their ACE2 receptors. Damage of pericytes (follow infection) could result in the induction of inflammation, cytokines storm, and thrombosis (aggregation of platelets, fibrin deposition, increase in D-Dimer, VWF protein, and factor VIII) which all reported in severe cases of COVID-19 (Becker, 2020; He et al., 2020; Helms et al., 2020). Although, most diagnostic kits for SARS-CoV-2 depend on naso- and oro-pharyngeal swabs, the viral RNA was also detected in blood and was used as a predictor for patient severity case (Chen et al., 2020b). Although SARS-CoV-2 can invade and replicate in human microvascular organoids containing pericytes (Monteil et al., 2020), further experiments are required to prove the ability of SARS-CoV-2 to infect pericytes.

In the heart, a higher expression for ACE2 was also found in pericytes as compared to other cellular subtypes (Chen et al., 2020a). In the lung, some complicated cases of COVID-19 suffered from fibrosis (Wang et al., 2020a). One of the main cellular sources for the developed myofibroblasts is the pericytes (Hung et al., 2013). Interestingly, pericytes expressed not only ACE2 but also CD147 receptors (Ulrich and Pillat, 2020), suggesting these cells as a more appropriate target for SARS-CoV-2.

In the kidney, pericytes located around either afferent arteriole of glomeruli or capillaries between renal tubules and could give rise to podocytes, mesangial cells, capillary smooth muscles, and fibroblast (Kurtz, 2016). Glomerular pericytes secrete renin (which acts as barosensor) and interstitial pericytes secrete erythropoietin (which acts as an oxygen sensor) (Stefanska et al., 2016). Both renin and erythropoietin play an important role in the maintenance of blood pressure and volume. Interestingly, renal pericytes also expressed ACE2 receptors, and severe cases of COVID-19 showed clear nephropathy with a significant reduction of renin, aldosterone, and erythropoietin (Rabb, 2020). I speculated that SARS-CoV-2 might infect renal pericytes causing their damage which could lead to decrease renin, aldosterone, and erythropoietin serum level and subsequently increase vascular leakage and hypoxia. This increased permeability and leakage could also occur due to lower ACE2 activity that indirectly induces the kallikrein–bradykinin pathway (Teuwen et al., 2020). Surprisingly, furin and its downstream target membrane type-1 metalloprotease (MT1-MMP) are over-expressed in podocytes, specialized form of pericytes, in diabetic rats (Boucher et al., 2006). MMPs are also activated by CD147 (Ulrich and Pillat, 2020). However, it is unknown whether CD147 and furin are co-expressed in pericytes.

It is worth to say that vascular damage precedes lung damage in COVID-19 patients (Becker, 2020; Helms et al., 2020). In support with early vascular damage, most or even all infected cases showed symptoms of loss of smell and taste and this was attributed to abundant co-expression of ACE2 and TMPRSS2 in pericytes in the olfactory bulb as well as basal and sustentacular cells in the olfactory mucosa, suggesting a higher infection rate in these cells (Brann et al., 2020). For this important role of pericytes in COVID-19 pathogenesis and the early vascular damage, I suggest an alternative name for the virus which is severe acute vascular and respiratory syndrome CoV (SAVRS-CoV).

**EV piece**

SARS-CoV-2 can escape from the initial host immune response (as indicated by delayed type I interferon production), explaining the presence of asymptomatic carriers (Prompetchara et al., 2020). SARS-CoV-2 could likely use other routes to be transmitted from cell to cell and to evade from the immune response of infected cells. Extracellular vesicles (EVs), including microvesicles and exosomes, which play an important role in transmission and pathogenesis of some viruses such as HBV, HCV, and respiratory syncytial virus (RSV) (Bukong et al., 2014; Chahar et al., 2018; Devhare et al., 2017; Muus et al., 2020), could be one of these routes which may be involved in SARS-Cov-2 transmission from cell to cell. In HCV, the virus escapes from host cell innate immune response through encapsulating their dsRNAs (a viral IFN inducer) within EVs (Grünvogel et al., 2018). The delayed secretion of type I interferon in the case of COVID-19 could theoretically occur in the same way, but this needs further investigations to practically prove this theory.

Biogenesis of EVs shares some cellular components (such as endosomes, multivesicular bodies and lysosomes) that hijacked by virus after entrance into cell (Chahar et al., 2018). EVs secreted from A549 lung epithelial cells infected with lentivirus loaded with some SARS-CoV-2 genes were uptaken by human-induced pluripotent stem cell-derived cardiomyocytes (hiPSC-CMs) (Kwon et al., 2020). The viral genes also induced the expression of inflammation-related genes in hiPSC-CMs. Moreover, exosomes derived from a cell culture system and patient's sera with complete replication of HCV genome showed an ability to transmit infection to normal human hepatocytes (Bukong et al., 2014; Lindenbach et al., 2005). Exosomes can activate immune response through induction of cytokine release during cytokine storm associated with RSV infection (Chahar et al., 2018). Interestingly, the exosomes isolated from the blood of hepatitis B patients contained protein and nucleic acids of HBV and exosomal contents were also detected in the natural killer cells and contributed to the suppression of immunity (Yang et al., 2017).

**Conclusions**

Every day we hear or read new information regarding SARS-CoV-2, but the only true thing is that there is no appropriate treatment for COVID-19, and developing an effective vaccine could take a long time. This virus resembles a Frankenstein monster as it has many strange pieces giving it unique properties over those of previous coronaviruses. In this review I discussed only 5 pieces, however, many other pieces remained mysterious (Fig. 1). Our question is which hand assembled these pieces? Is the hand of a scientist, God, or nature? Who wants to punish the world? Is this Frankenstein monster assembled in the laboratory during a vaccine development trial on bats and was disseminated unintendedly to people? Is this a message from God to follow his guidelines for peace, safety, fairness, and justice? Is this a message from nature to decrease pollution and maintain the ecosystem and green area? We have no answer at this time, however, the next days could have suitable answers and then all monster pieces could be discovered.

**Acknowledgment**

I would like to thank the medical illustrator Ashraf Ragab Refaey (a student in the faculty of Veterinary medicine, Kafrelsheikh University) for the hand sketching SARS-CoV-2 Frankenstein monster.

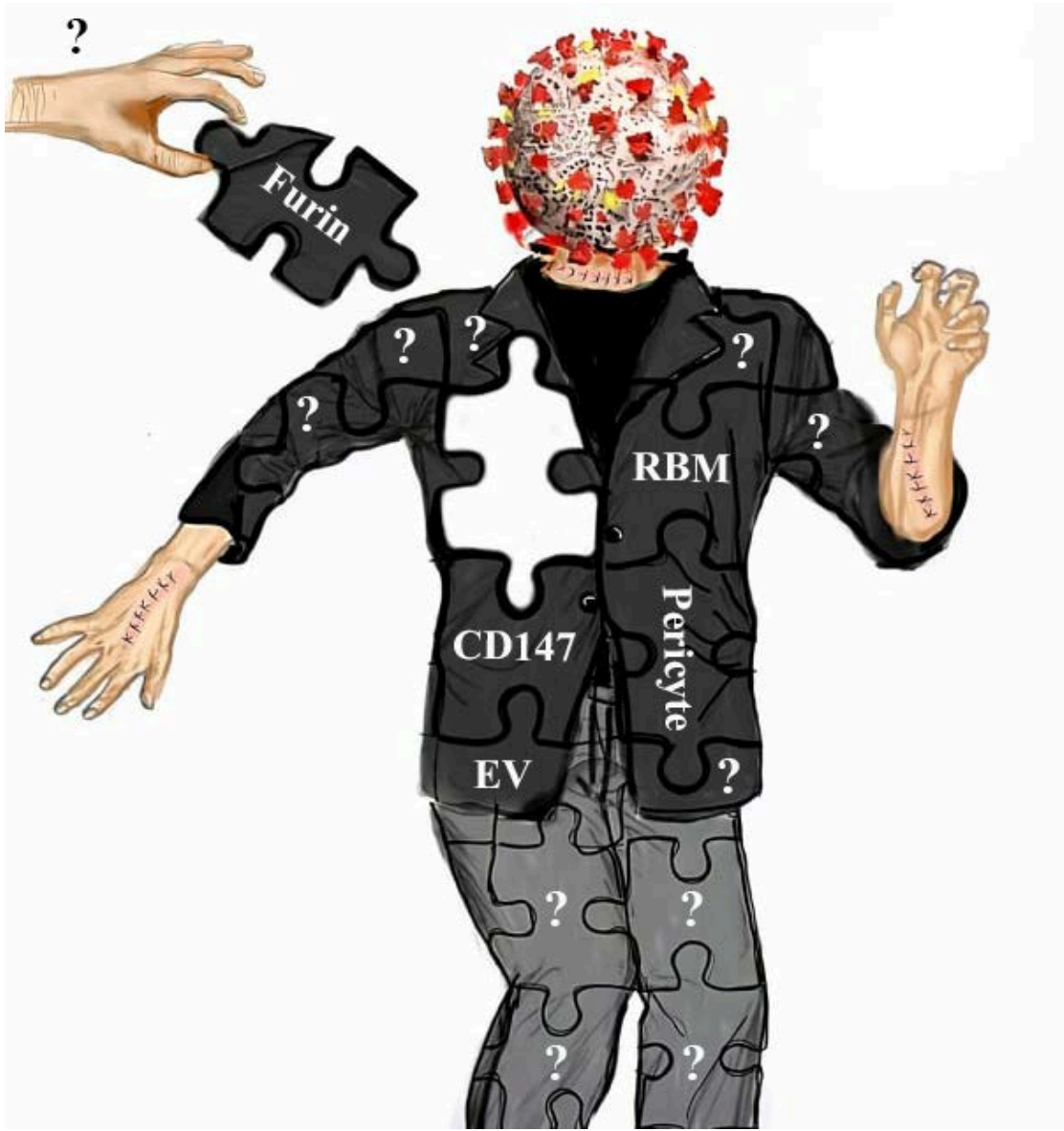

Fig.1. An unknown hand assembled the SARS-CoV-2 Frankenstein monster. Only five pieces of the monster were known, but many remained mysterious.